\documentclass[12pt]{article}
\usepackage[dvips]{graphicx}

\usepackage{amssymb}
\usepackage{amsmath}
\usepackage{latexsym}
\usepackage{subfigure}
\usepackage{color}

\def\lromn#1{\uppercase\expandafter{\romannumeral#1}}

\def\gsim{\:\raisebox{-0.5ex}{$\stackrel{\textstyle>}{\sim}$}\:}

\begin{document}

\begin{titlepage}
  \begin{center}

    \hfill KEK Preprint 2010-41 \\
    \hfill FTUV-10-1129\\
    \hfill \today

    \vspace{0.4cm}
    {\large\bf Measuring Anomalous Couplings in $H\to W W^\ast$ Decays at the International Linear Collider}
    \vspace{0.8cm}

    {\bf Yosuke Takubo}$^{(a)}$,
    {\bf Robert N. Hodgkinson}$^{(b)}$,
    {\bf Katsumasa Ikematsu}$^{(a)}$,\\
    {\bf Keisuke Fujii}$^{(a)}$, 
    {\bf Nobuchika Okada}$^{(c)}$, and 
    {\bf Hitoshi Yamamoto}$^{(d)}$

    \vspace{0.8cm}

    {\it
     $^{(a)}${High Energy Accelerator Research Organization (KEK), Tsukuba, Japan} \\
     $^{(b)}${Departamento de F\`isica Te\`orica and IFIC, Universitat de Val\`encia--CSIC, Val\`encia, Spain} \\
     $^{(c)}${Department of Physics and Astronomy, University of Alabama, Tuscaloosa, AL 35487, USA} \\
     $^{(d)}${Department of Physics, Tohoku University, Sendai, Japan} \\
     }

    \vspace{0.8cm}

    \abstract{Measurement of the Higgs coupling to $W$-bosons is an important test of our understanding of the electroweak symmetry breaking mechanism.  We study the sensitivity of the International Linear Collider (ILC) to the presence of anomalous $HW^+W^-$ couplings using $ZH\to \nu\bar\nu WW^\ast\to \nu\bar\nu 4j$ events.  Using an effective Lagrangian approach, we calculate the differential decay rates of the Higgs boson including the effects of new dimension-$5$ operators.  We present a Monte Carlo simulation of events at the ILC, using a full detector simulation based on geant4 and a real event reconstruction chain.  Expected constraints on the anomalous couplings are given.}

  \end{center}
\end{titlepage}

\setcounter{footnote}{0}


\section{Introduction} 
\label{sec:intro}
The announcement of the discovery of the Higgs boson candidate at the Large Hadron Collider (LHC), has elevated the question of its properties to the top of the list of questions in high energy physics \cite{atlas, cms}. The Higgs field is crucial to the Standard
Model (SM), as it provides the mechanism for electroweak (EW) symmetry breaking and gives rise to the masses of both the $W^\pm$ and $Z$ gauge bosons and all charged fermions when it develops a vacuum expectation value (VEV).

In practice, the discovery of a SM-like Higgs boson is only the beginning of the story.  Over the last thirty years, a number of alternative models have been proposed, notably low-energy supersymmetry and models with extra dimensions, which attempt to resolve the so-called hierarchy problem of the SM.  It is crucial to measure the properties of the Higgs particle in order to ascertain to which, if any, of the proposed models it belongs.  Of particular importance will be the task of verifying that the discovered Higgs boson candidate really is the particle responsible for EW symmetry breaking and the generation of fermion masses.

Indeed, it is notable that one of the ``gold plated'' Higgs boson discovery channels at the LHC involves production of the Higgs particle through gluon fusion, followed by its decay into photon pairs, i.e. $gg\to H\to \gamma \gamma$.  This channel involves loop-induced couplings in both production and decay and the discovery of such a resonance tells us very little about EW symmetry breaking!\footnote{
In fact, a scalar in a certain class of new physics models \cite{IOY}, 
 which is nothing to do with the EW symmetry breaking, 
 can mimic such a signal. 
It has been studied \cite{FIOY} how well 
 the International Linear Collider distinguishes 
 the scalar from the Higgs boson. 
}  
For this, accurate measurements of the Higgs boson couplings to the EW gauge bosons must be made.
The LHC can be used to extract some information on the Higgs boson coupling to $Z$ bosons using
the decay $H\to Z Z\to (l^+l^-) (l'^+l'^-)$ \cite{Godbole:2007cn}, since this final state can be
efficiently triggered; effects of anomalous $HW^+W^-$ couplings can also be probed through their
contribution to $WW$ scattering \cite{Zhang:2003it} and the gauge-boson fusion Higgs production
mechanism \cite{Plehn:2001nj}.  
For a direct measurement of the $HW^+W^-$ coupling, however, the best
environment is
a lepton collider, such as the International Linear Collider (ILC).  

In an electron-positron collider
such as the ILC,
Higgs bosons are predominantly produced through
EW interactions, either through Higgs-strahlung from virtual $Z$-bosons or through gauge boson
fusion; such reactions can therefore be used to probe anomalous Higgs-gauge-gauge couplings
\cite{Hagiwara:1993sw}.  Furthermore, the clean environment of a lepton collider also allows for
studies based on the asymmetries in the decays of the Higgs boson \cite{Biswal:2005fh}.  For an additional heavy Higgs boson (with mass $\gsim 200$~GeV), if any, the authors of \cite{Niezurawski:2004ga} have
proposed to accurately measure the couplings instead at a future photon collider.  

If, as we expect, the underlying theory is gauge invariant, then anomalous $HWW$ couplings necessarily imply anomalous contributions to the $HZZ,\ H\gamma\gamma$, and mixed $HZ\gamma$ vertices.  Whilst measurements of anomalous Higgs boson couplings to the neutral vector bosons may make use of the very high rates of the Higgs-strahlung process, determining the structure of the $H W^+ W^-$ vertex relies on either the gauge-boson fusion production process or on the decay $H\to WW^\ast$.
Furthermore, these measurements should be performed independently of one another, as a test of the underlying gauge invariance.  In this paper we concentrate on the decay process for two reasons; firstly, the differential cross sections considered do not depend on the additional $HNN,\ N=Z,\gamma$ anomalous couplings at leading order and secondly, this process allows us to study the effects of CP-violating parameters.  In contrast, since the final state forward neutrinos are not measured in the gauge-boson fusion process $e^+e^-\to H \nu\bar\nu$, this process has no sensitivity to CP-violating parameters; furthermore, there are large backgrounds from the related process $e^+e^-\to H e^+e^-$, where the final state electrons also escape detection, which introduces a significant dependence on the $HNN$ anomalous couplings.


In particular, we study the feasibility of measurements for anomalous Higgs boson couplings to $W^+ W^-$ pairs using $ZH$ production followed by $H \to WW^{\ast}$ and $Z \to \nu \nu$ at the ILC, based on realistic Monte Carlo simulations. We stress that this is the first full simulation study of its kind using a full detector simulator based on geant4 and a real event reconstruction chain.  The Higgs production mechanism clearly depends on the presence of anomalous $HZZ$ and $HZ\gamma$ couplings, however the distributions of the Higgs boson decay products which constitute our signal do not.  The only possible effect of anomalous $HNN$ couplings in the production mechanism is a change in the overall rate $e^+e^-\to ZH \to ZWW^\ast$ and can easily be measured at relatively low integrated luminosities \cite{Hagiwara:1993sw}. Hence, we feel justified in neglecting the effect of these couplings in our analysis focusing on the Higgs decay $H \to WW^\ast$.

The structure of our paper is as follows.  
In the next section we outline the effective interaction Lagrangian
under discussion and present analytic formulas for the relevant 
differential decay rates.  
In Section~\ref{sec:simulation}, 
we present details of the Monte Carlo simulation 
and results of the analysis. 
We discuss these results and a simple example of 
new physics model which give rise to the effective 
interaction Lagrangian in Section~\ref{sec:discussion}. 
The final section is reserved for summary and conclusions.

\section{Physics Model}
\label{sec:Model}
We may parametrise the relevant terms of the general interaction Lagrangian, which couples the Higgs boson to EW vector bosons in a Lorentz-symmetric fashion, as
\begin{equation}
\label{eqn2.1}
\mathcal{L}_{\rm HWW}
=
2 M_W^2\left(\frac{1}{v}+\frac{a}{\Lambda}\right) H\ W_\mu^+ W^{-\mu}
+\frac{b}{\Lambda} H\ W^+_{\mu\nu} W^{-\mu\nu}
+\frac{\tilde b}{\Lambda} H\ \epsilon^{\mu\nu\sigma\tau} W^+_{\mu\nu} W^-_{\sigma\tau}\ ,
\end{equation}
where $M_W$ is the mass of the $W$-boson, $W_{\mu\nu}^\pm$ is the usual gauge field strength tensor,
$\epsilon^{\mu\nu\sigma\tau}$ is the Levi-Civita tensor, $v$ is the VEV of the Higgs field, $a,b,\tilde{b}$
are real dimensionless coefficients and $\Lambda$ is a cutoff scale.  The SM interaction is recovered in the limit $a,b,\tilde{b}\to 0$.  The dimensionless couplings $b, \tilde{b}$ parametrise the
leading dimension-five non-renormalisable interactions\footnote{The effects of dimension-six operators in
the effective Lagrangian were considered in~\cite{Dutta:2008bh}.}, which we assume are due to contributions
arising from some new physics at the scale $\Lambda$.  
The dimensionless coupling $a$ represents corrections
to the SM term, assumed to originate at the same scale $\Lambda$.
The Lagrangian (\ref{eqn2.1}) is not by itself gauge invarient; to restore explicit gauge invarience we must also include the corresponding anomalous couplings of the Higgs boson to $Z$ bosons and photons.

We will assume the Higgs boson mass to be $M_H<2 M_W$, being consistent with the recent discovery of the Higgs boson candidate at the LHC \cite{atlas, cms}, so that the decay to real $W^+ W^-$ pairs is
kinematically forbidden; the anomalous couplings may however contribute to the decay $H\to W W^\ast$ with
distinct signatures.  The $a$ parameter is simply a rescaling of the SM coupling and therefore manifests itself
as a shift in the overall partial width for this channel.  By comparison, the non-renormalisable coupling $b$
has a different Lorentz structure to the SM term and leads to a change in the ratio of couplings to the transverse
or longitudinal components of the gauge bosons.  Finally, the coupling $\tilde b$ introduces a CP-violating
operator which can affect angular correlations, as discussed below.

\begin{figure}
\begin{center}
\includegraphics[width=0.48\columnwidth]{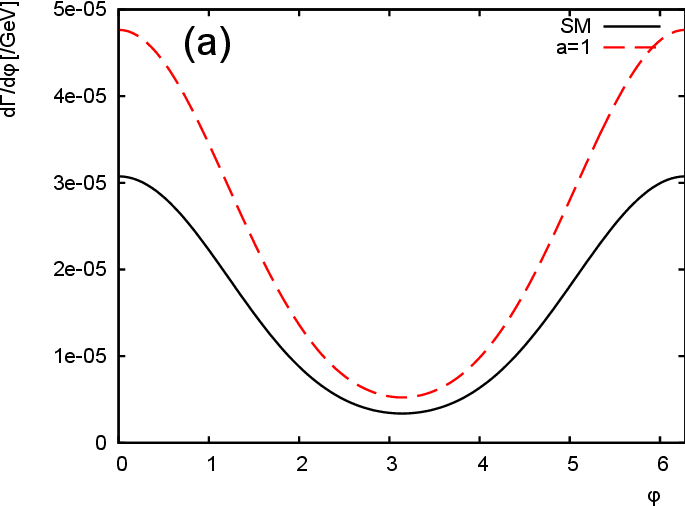}
\includegraphics[width=0.48\columnwidth]{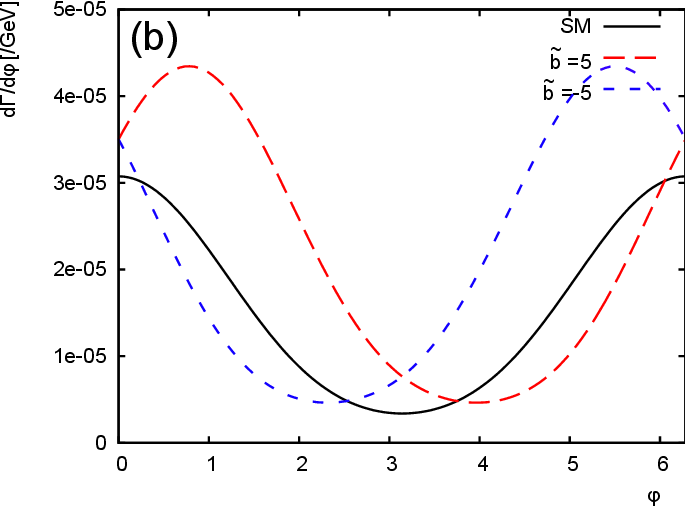}
\caption{The $\phi$-distribution of the decay $H\to WW^\ast\to 4j$ with the inclusion of anomalous couplings.
(a) The SM curve along with that for $a=1$, $b=\tilde{b}=0$, 
same for both distributions.  (b) The SM result with the cases $\tilde{b}=\pm5$, $a=\tilde{b}=0$, $\Lambda=1$~TeV;
the position of the minimum is now shifted as discussed in the text.}
\label{Fig:phi}
\end{center}
\end{figure}


Assuming all final state fermions to be massless, the differential partial width for the decay chain
$H\to W W^\ast\to 4j$ as a function of the on-shell $W$-boson momentum $p_W$ and the azimuthal angle between the
up-type quark and anti-quark $\phi$ (with axis of rotation in the direction of the $W^+$ momentum) is given by
\begin{eqnarray}
\label{decayrate}
\frac{d^2\Gamma}{d p_W d \phi}
& = &
\left(N_c \sum_{\stackrel{i=u,c}{j=d,s,b}} |V_{ij}|^2\right)^2
\frac{1}{2M_H\ 2M_W\Gamma_W}
\frac{ M_W^{10} k^2 p_W^2}{(k^2-M_W^2)^2 \pi^5 E_W v^6}
\nonumber\\
& & \hspace{1cm} \times
\left[\frac{c_0}{36}
+\left(\frac{c_{\phi} e^{i\phi}+c_{\phi}^\ast e^{-i\phi}}{512}\right)
+\left(\frac{c_{2\phi} e^{i2\phi}+c_{2\phi}^\ast e^{-i2\phi}}{288}\right)
\right]\ ,
\end{eqnarray}
where $N_c=3$ is the number of colors, $V_{ij}$ is the Cabibbo-Kobayashi-Maskawa quark mixing matrix,
$\Gamma_W$ is the $W$-boson width, $E_W=\sqrt{M_W^2+p_W^2}$ is the energy of the on-shell $W$-boson and
$k^2=(M_H-E_W)^2-p_W^2$ is the invariant squared mass of the off-shell $W$-boson.  The coefficient functions
$c_0,\ c_\phi,\ c_{2\phi}$ can be written in terms of two dimensionless combinations of parameters,
\begin{eqnarray}
\label{phicoefficients}
c_0 & = & |T|^2+\frac{1}{2} L^2\ ,\nonumber\\
c_\phi & = & \pi^2 TL\ ,\nonumber\\
c_{2\phi} & = & T^2\ ,
\end{eqnarray}
where the real function $L$ is due to contribution from longitudinally polarized on-shell $W$-bosons and
the complex function $T$ from transversely polarized bosons.  Explicitly, they are given by
\begin{eqnarray}
T & = & \left(1+\frac{a v}{\Lambda}\right)
-\frac{b v}{\Lambda} \frac{\left(p^\mu_W k_\mu\right)}{M_W^2} 
-i \frac{2M_H p_W}{M_W^2}\frac{v\; \tilde{b}}{\Lambda}\ ,
\nonumber\\
L & = & \left(1+\frac{a v}{\Lambda}\right)
\frac{\left(p^\mu_W k_\mu\right)}{M_W k}
-\frac{b v}{\Lambda} \frac{\left(\left(p^\mu_W k_\mu\right)^2-M_H^2 p_W^2\right)}{M_W^3 k}\ ,
\end{eqnarray}
where $k_\mu$ is the $4$-momentum of the off-shell $W$-boson and $k=\sqrt{k_\mu k^\mu}$.  The 4-vector
product $p^\mu_W k_\mu$ can be expanded as $p^\mu_W k_\mu=M_H E_W-M_W^2$.

\begin{figure}
\begin{center}
\includegraphics[width=0.48\columnwidth]{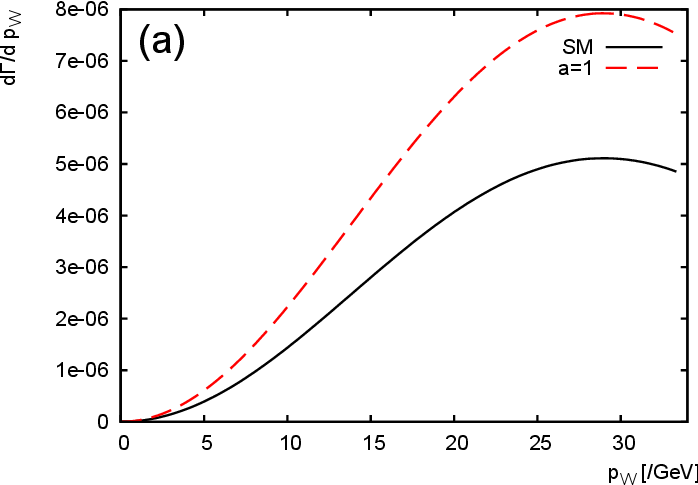}
\includegraphics[width=0.48\columnwidth]{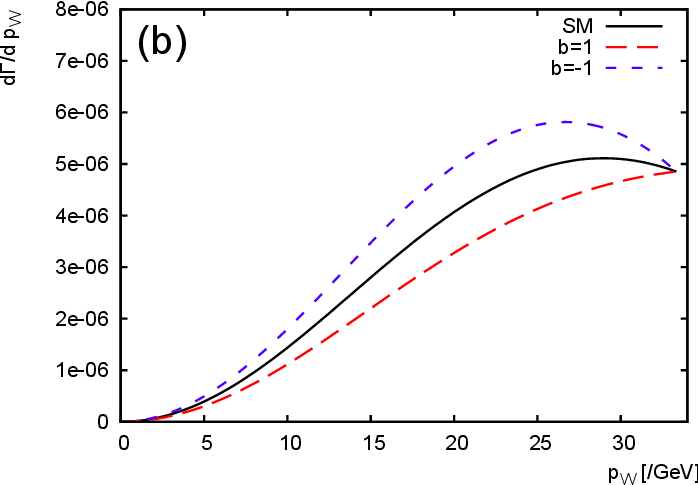}
\caption{The $p_W$ distribution of the decay $H\to WW^\ast\to 4j$ with the inclusion of anomalous couplings. (a) The SM curve along with that for $a=1$, $b=\tilde{b}=0$, $\Lambda=1$~TeV. (b) The SM result with the cases $b=\pm 1$, $a=\tilde{b}=0$, $\Lambda=1$~TeV.}
\label{Fig:pw}
\end{center}
\end{figure}


We see that for $\tilde{b}=0$, all coefficients $c_0,\ c_\phi,\ c_{2\phi}$ are real and the partial
decay width is a function of cosines only.  In the SM limit, the magnitudes of the transverse
function $T$ and the longitudinal function $L$ are approximately equal, assuming a $W$-boson energy
of the order of the $W$ mass, $E_W\sim M_W$.  From (\ref{phicoefficients}), in this limit we have the
ratio of coefficients $c_\phi/c_{2\phi}\sim \pi^2$ so that the $\cos\phi$ term dominates, the minimum
of the distribution is seen to be at $\phi_{\rm min} =\pi$.  Non-zero values of $\tilde{b}$ shift the
minimum of the $\phi$ distribution to $\phi_{\rm min} =\pi+\delta$
with $\delta={\rm arg}(T)$.  
This effect is illustrated in Fig.~\ref{Fig:phi}, 
where we plot the $\phi$-dependence of the partial width 
in both the SM and taking 
 $a=1, b=\tilde{b}=0$ (Fig.~\ref{Fig:phi} (a)) 
 and $a=0, \tilde{b}=\pm5$ (Fig.~\ref{Fig:phi} (b)), 
 with $\Lambda=1$~TeV.

The energy of the on-shell $W$ boson in the decay $H\to WW^\ast$ is not fixed by kinematic
constraints.  After integrating (\ref{decayrate}) over $\phi$ we see that only the $c_0$ term
contributes to the differential decay rate $d\Gamma/d p_W$.  The presence of the anomalous couplings
$b,\tilde{b}$ modifies the energy-dependence of this expression through the $L$ and $T$ functions, as
shown in Fig.~\ref{Fig:pw}, which compares the effect of non-zero $a$ and $b$ terms.  In particular,
the contribution of $b$ is seen to vanish at the kinematic limit of the distribution.

\section{Monte Carlo Simulation} \label{sec:simulation}
\subsection{Simulation Conditions and Tools}\label{sec:mctool}
For the simulation of the measurement of the anomalous $HWW$ couplings, we take the Higgs mass to be $120\,$GeV along with a center of mass energy of $250\,$GeV and an integrated luminosity of $250\,$fb$^{-1}$ as assumed in the Letters of  Intent (LoI) for the ILC detectors \footnote{For historical reasons, the Higgs mass assumed in the most Higgs related studies for the ILC have been performed with the Higgs mass of $120\,$GeV. Since the update of the Higgs mass to $125\,$GeV takes long time for full simulation studies like that presented in this paper, we use the $120\,$GeV in this paper. The modification of the Higgs mass to $125\,$GeV will not alter the basic conclusions from this analysis in any qualitative manner, though it might alter the sensitivity to the anomalous couplings slightly due to the increase of the $H \to WW^{\ast}$ branching ratio.}\cite{ild, sid, 4th}.
Notice that for a Higgs mass of $120\,$GeV the $ZH$ cross-section attains its maximum at around this energy\footnote{
Note also that the mass resolution for the Higgs boson recoiling against lepton pairs from the accompanying
$Z$ boson is known to degrade with the energy.  The most accurate determination of the Higgs mass is hence expected
in this energy region, by investing a substantial running time to accumulate an integrated luminosity of
$250\,$fb$^{-1}$, together with a model-independent determination of the total $ZH$ production cross section,
which is indispensable to measurements of the various branching ratios of the Higgs boson.
}
and the branching ratio for $H \to WW^{*}$ decay is 15.0\%, which is subdominant.  The angular analyses and
$W$ momentum measurement discussed in the previous section necessitate the identification of the four jets from
the $H \to WW^{\ast}$ decay and their correct pairing.  Given the branching fraction of this decay mode and the
additional combinatorial background in the jet paring which would otherwise hamper the analyses, we require the
associated $Z$ to decay into $\nu \bar{\nu}$.  Our signal thus consists of four jets plus missing energy.
Consequently, four fermion final states such as $\nu \nu qq$, $qq \ell \nu$, $\ell \ell \ell \ell$, 
$qq \ell \ell$, and $qqqq$, primarily coming from $WW$ and $ZZ$ production, comprise the SM background. In order
to suppress these backgrounds, which would otherwise dominate in this study, we use 80\% right-handed polarization
for the electron beam and 30\% left-handed polarization for the positron beam in the following analysis.

The signal events $ZH \to \nu \nu WW^{\ast}$ were generated using Physsim \cite{physsim}, with the cutoff scale
$\Lambda$ taken to be 1 TeV when the anomalous couplings were switched on\footnote{
Note that the absolute values of the anomalous couplings such as $a$, $b$, and $\tilde{b}$ become meaningful only
after the cutoff scale is given.
}.
$ZH \to \nu \nu H$ events for decay modes other than $H \to WW^{\ast}$ were generated by WHIZARD, together with
all the other SM backgrounds.  In both generators, initial-state radiation and beamstrahlung have been included 
in the event generation.  The beam energy spread was set to 0.28\% for the electron beam and 0.18\% for the
positron beam.  We have ignored the finite crossing angle between the electron and positron beams. 
In the event generation, helicity amplitudes were calculated using the HELAS library \cite{helas}, which allows
us to deal with the effect of gauge boson polarizations properly. 
The event generator Pythia6.409 \cite{pythia} was used for parton-showering and hadronization.  The generator data for the SM background
had been prepared as a common data sample for the LoI studies and stored in the StdHep format \cite{stdhep} at SLAC.
The SM background sample consists of all the SM processes with up to 4 fermions in the final state, which is about
10 million events in total. Since the cross-section of 6 fermion events is small compared to that of the signal at
$\sqrt{s} = 250$ GeV and can be rejected easily, we ignored these events.

The 4-momenta of the (quasi-)stable particles after parton-showering and hadronization were fed into a geant4-based
full detector simulator called Mokka \cite{mokka}, in which \verb|ILD_00| is implemented as the detector model \cite{ild}. 
The \verb|ILD_00| detector model consists, from inside to outside, of a very thin 6-layer vertex detector with a point
resolution of about $3\,\mu$m, silicon internal and forward trackers, a time projection chamber (TPC) having about 200
sample points with a point resolution of $100\,\mu$m or better, silicon external and endcap trackers, ultra high granularity
electro-magnetic and hadron calorimeters, a superconducting solenoid of $3.5\,$T, and return yokes interleaved with muon
detectors.  With this detector model, we  expect the transverse momentum resolution 
($\Delta(1/p_{\mathrm{T}}) = \Delta p_{\mathrm{T}}/p_{\mathrm{T}}^{2}$) 
to be $2.0 \times 10^{-5}$ GeV$^{-1}$ asymptotically, rising to $9.0 \times 10^{-5}$ GeV$^{-1}$ at 10 GeV, 
and to $9.0 \times 10^{-4}$ GeV$^{-1}$ at 1 GeV. 

The generated detector hits and signals were processed through a real event reconstruction program called MarlinReco
implemented in the Marlin framework \cite{marlin}.  In the event reconstruction, charged particle tracks were reconstructed
from tracker hits by a realistic track finder and a Kalman-filter-based track fitter, taking into account signal overlapping
as well as energy loss and multiple scattering.  Calorimeter hits were then clustered and combined with the tracker
information to perform a particle flow analysis (PFA) \cite{pfa} to achieve the best jet energy resolution.  The jet energy
resolution for $45\,$GeV jets from $Z \to q\bar{q}$ events is estimated to be 3.7\%,  which improves to about 3\% for $100\,$GeV
jets.  Jet clustering was done with the Durham algorithm \cite{durham}, and the resultant jets were flavor-tagged on a
jet-by-jet basis with the LCFIVertex package \cite{lcfi} after vertex finding with the ZVTOP algorithm \cite{zvtop}.

\subsection{Event Selection} \label{sec:selection}

The goal of our event selection is to isolate the signal events with four jets plus missing energy
originating from $ZH$ production followed by $H \to WW^{\ast}$ and $Z \to \nu\bar{\nu}$ decays. We thus
started our event selection by forcing all the events to cluster into four jets by adjusting the
$Y_{\rm cut}$ value  \cite{durham}.
The Higgs boson and on-shell $W$ boson masses were then reconstructed by paring these four jets so as to
minimize the $\chi^{2}$ function defined by
\begin{equation}
\chi^{2} = 
\frac{(^{\mathrm{rec}}M_{H} - M_{H})^{2}}{\sigma_{H}^{2}} +
\frac{(^{\mathrm{rec}}M_{W} - M_{W})^{2}}{\sigma_{W}^{2}},
\end{equation}
where $^{\mathrm{rec}}M_{H}$ is the reconstructed Higgs mass, $M_{H}$ is the input
Higgs mass (120 GeV), $^{\mathrm{rec}}M_{W}$ is the reconstructed on-shell $W$ mass, $M_{W}$ is the nominal
$W$ mass ($80.4\,$GeV) and $\sigma_{H(W)}$ is the mass resolution for the Higgs@($W$). 

After the mass reconstruction, we required the reconstructed Higgs mass ($^{\mathrm{rec}}M_{H}$) 
to lie in the range $110\,\mathrm{GeV} <\, ^{\mathrm{rec}}M_{H} < 130\,\mathrm{GeV}$.  Since we assume
a $Z$ boson decaying into a neutrino pair in the signal event, resulting in a missing mass peak at the $Z$ mass, 
we required a missing mass in the range $70\,$GeV $<\, ^{\mathrm{miss}}M < 140\,$ GeV. 

The main backgrounds in this analysis are from $e^+e^- \to WW$ and $ZZ$.  The angular distributions of these
processes have peaks in the forward and backward regions.  For this reason, we required the angle of the
reconstructed Higgs boson with respect to the beam axis ($\cos \theta_{H}$) to be
$|\cos \theta_{H}| < 0.95$.  We then looked at the $Y$-value for the forced 4-jet clustering, which is
expected to be small for $\nu \nu qq$ and $\nu \nu \ell \ell$ events
having only two ``partons'' in their final states.  We therefore selected events with $Y_{\mathrm{-}}>0.0005$, 
where $Y_{\mathrm{-}}$ is the threshold $Y$-value at which the number of jets changes four to three.  After the
selection cuts described so far, the dominant background became $\ell \nu qq$.  The lepton in the $\ell \nu qq$
final state comes from the leptonic decay of a $W$ and has a larger energy than leptons from jets.  We hence
required the maximum track energy ($E_{\mathrm{trk}}$) to be below 30 GeV. 

Since we are focusing our attention on the  $H \to WW^{\ast}$ decay in this analysis, the $ZH \to \nu \nu bb$
process is a background to be discarded.  We thus rejected $ZH \to \nu \nu bb$ events by requiring the number
of $b$-tagged jets ($^{\mathrm{4-jet}}N_{b}$) to be $^{\mathrm{4-jet}}N_{b} \leq 1$.  Since the
$ZH \to \nu \nu bb$ channel has two jets in the final state, candidate events were further jet-clustered into
two jets and then required to have no $b$-tagged jets, $^{\mathrm{2-jet}}N_{b} = 0$.

\begin{figure}
\centerline{\includegraphics[width=0.43\columnwidth]{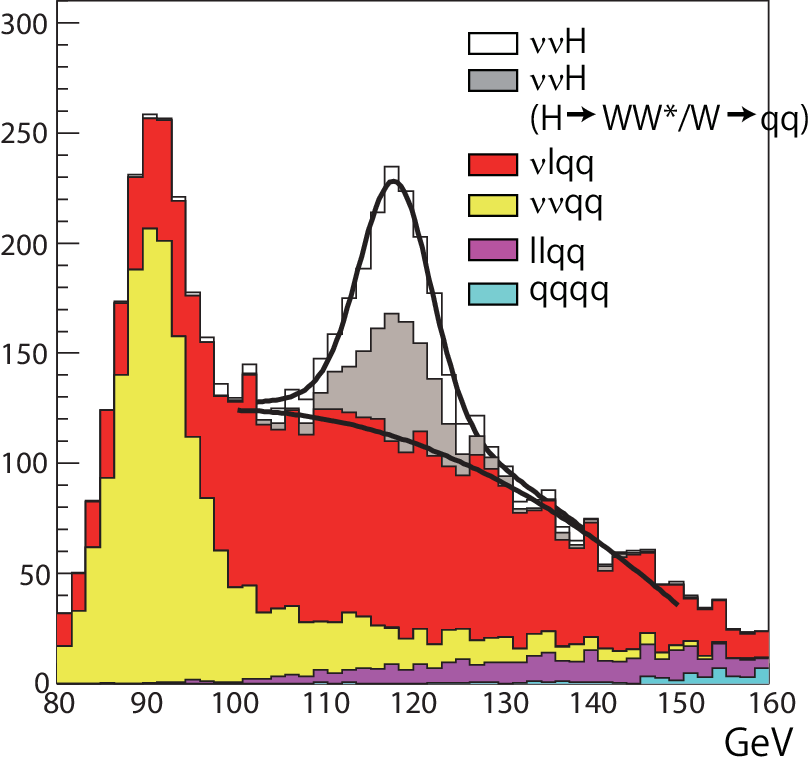}}
\caption{Distribution of the reconstructed Higgs mass after selection cuts.}
\label{fig:hmass}
\end{figure}

After all the selection cuts, we performed a likelihood analysis as follows.  We used $^{\mathrm{miss}}M$,
$\cos \theta_{H}$, $Y_{-}$, $^{\mathrm{4-jet}}N_{b}$, and the number of charged tracks as the input
variables of the likelihood function and tuned the likelihood cut position to maximize signal significance. 
We obtained a maximum signal significance of $7.6$ at a likelihood cut position $\mathcal{L}_{\mathrm{cut}}= 0.79$. 
Figure \ref{fig:hmass} shows the reconstructed Higgs mass distribution after all cuts.  Fitting the distribution
with a double Gaussian plus a second order polynomial, we estimated the expected accuracy of the branching ratio
BR($H \to WW^{\ast}$) to be $15.7\%$,  assuming that the measurement accuracy of the $ZH$ cross-section is
$2.5\%$ \cite{ild}.  The branching ratio can be determined to an accuracy of $5\%$, however, by using processes
with leptonic decays of the $W$ \cite{rdr}.  The number of events before and after the selection cuts are
summarized in Table \ref{tb:cut_summary}.

\begin{table}
\center{
\begin{tabular}{|l|r|r|r|r|}
\hline
Process  & No cut & After cuts & $\mathcal{L}_{\mathrm{cut}} >0.79$ & $N_{c} = 2$ \\
\hline
$\nu \nu H (H \to \mathrm{all})$ & 10,634 & 1,518 & 756 & 546 \\
$\nu \nu H (H \to WW^{\ast} \to $ 4-jet) & 680 & 512 & 348 & 258 \\ \hline
$\ell \ell \ell \ell$ & 753,964 & 46 & 0 & 0 \\
$qqqq$ & 378,726 & 8 & 3 & 2 \\
$\ell \ell qq$ & 335,762 & 409 & 94 & 70 \\
$\nu \ell qq$ & 299,866 & 8,571 & 1,063 & 692 \\ 
$\nu \nu \ell \ell$ & 103,704 & 3 & 0 & 0 \\
$\nu \nu qq$ & 63,649 & 1,090 & 207 & 110 \\ \hline
\end{tabular}
}
\caption{Cut summary.}
 \label{tb:cut_summary}
\end{table}

\subsection{Analysis Results} \label{sec:analysis}
We investigated the distributions of the variables sensitive to the anomalous couplings.  The distributions of the $W$ boson momenta in the Higgs
rest-frame ($p_{W}$) and the jet angle in the $W$ boson rest-frame were plotted after selection cuts.  The jet angle distributions were plotted
for the on-shell ($\cos\theta_{j1}$) and off-shell $W$ bosons ($\cos\theta_{j2}$) separately.  We also examined the distribution of the angle
between the two $W$ boson decay planes ($\phi_{\mathrm{plane}}$) corresponding to that between the two up-type quarks from the decays of the $W$
bosons.  We applied double $c$-tagging to select the two up-type quarks ($c$ quark) in $ZH \to \nu \nu WW^{*} \to \nu\nu cscs$, where the
selection efficiency was 88\% as shown in Table \ref{tb:cut_summary}.  The $\phi_{\mathrm{plane}}$ was histogrammed using the two $c$-tagged
jets, without identifying their charges.  Distributions for the $ZH$ events were obtained, evaluating the contamination from the SM backgrounds
by fitting the Higgs mass distribution for each $\phi_{\mathrm{plane}}$ bin.  Since the branching ratio of the Higgs to channels other than
$H \to WW^{\ast}$ will be determined with much better accuracies than the statistical errors shown on the distributions \cite{rdr}, we
subtracted the background from these decay modes ignoring their systematic  errors on the cross-section to obtain the distributions of $p_{W}$, $\cos\theta_{j1}$, $\cos\theta_{j2}$, and $\phi_{\mathrm{plane}}$ for $ZH \to \nu \nu WW^{\ast}$ events. 

As mentioned above, if the anomalous couplings exist in $H \to WW^{\ast}$, there should also be similar anomalous couplings of the same origin in $H \to ZZ^{\ast}$ and $Z \gamma$ decays. In order to make sure that the possible anomalies in the $HZZ$ and $HZ\gamma$ couplings would not affect our measurement of the anomalous $HWW$ couplings, we have evaluated the contamination from the $H \to ZZ^{\ast}$ and $Z\gamma$ decays. After all the selection cuts, the contamination of the $H \to WW^{\ast}$ sample from the $H \to ZZ^{\ast}$ and $H \to Z\gamma$ decays was only 23 events in the SM case and within the statistical error.

As long as the anomalous couplings stem from the same origin, it is reasonable to expect that the anomalous $HZZ$ and $HZ\gamma$ couplings, are of the same order to the anomalous $HWW$ couplings. The effect of the anomalous $HZZ$ and $HZ\gamma$ couplings on the contamination from the $H \to ZZ^{\ast}$ and $H \to Z\gamma$ decays (only the 23 events in the SM) in the $H \to WW^{\ast}$ sample can be ignored, since, if the effect on the $H \to ZZ^{\ast}$ and $H \to Z\gamma$ contamination is sizable, the effect on the $H \to WW^{\ast}$ should be much larger for our $H \to WW^{\ast}$ sample as long as the interference term with the SM amplitude dominates the anomalous coupling term squared. We thus conclude that the possible anomalous $HZZ$ and $HZ\gamma$ couplings will not affect the sensitivity of our measurement of the anomalous $HWW$ couplings using the $H \to WW^{\ast}$ decay.

It should also be worth noting that we can separately study the effect including its size of the anomalous $HZZ$ and $HZ\gamma$ couplings, for instance by measuring the production cross section: $e^{+} e^{-} \to ZH$ without looking at the Higgs decay at all, using the recoil mass technique \cite{ild}.

\begin{figure}
\centerline{\includegraphics[width=12cm]{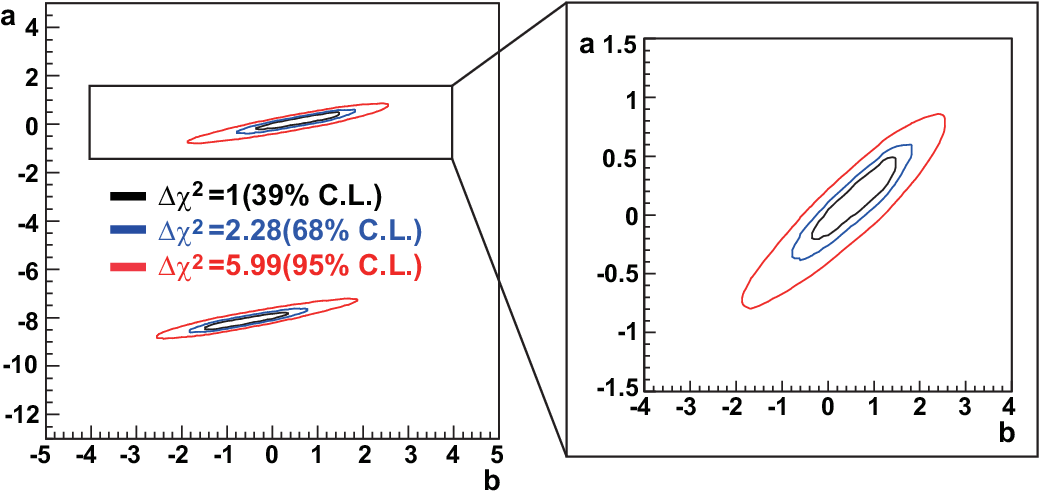}}
\caption{Probability contours for $\Delta \chi^{2} =$ 1, 2.28, and 5.99 in the $a$-$b$ plane, which correspond to 39\%, 68\%, and 95\% C.L., respectively.}
\label{fig:cont_a-b}
\end{figure}

\begin{figure}
\centerline{\includegraphics[width=12cm]{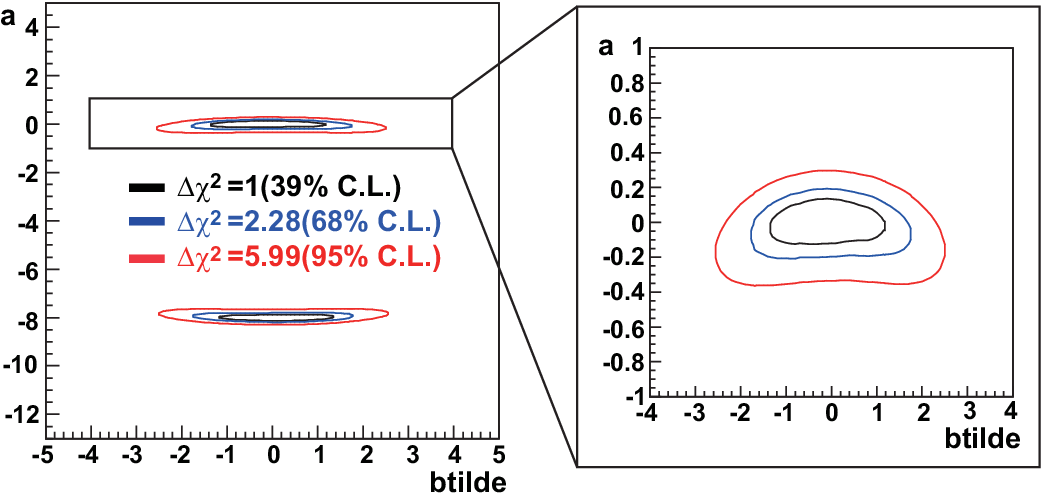}}
\caption{Contours similar to Fig.$\,$\ref{fig:cont_a-b} plotted in the $a$-$\tilde{b}$ plane.}
\label{fig:cont_a-bt}
\end{figure}

\begin{figure}
\centerline{\includegraphics[width=6cm]{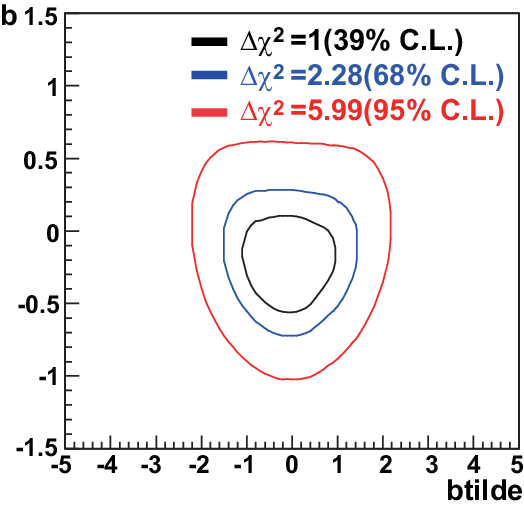}}
\caption{Contours similar to Fig.$\,$\ref{fig:cont_a-bt} plotted in the $b$-$\tilde{b}$ plane.}
\label{fig:cont_b-bt}
\end{figure}

To estimate the sensitivity to the Higgs anomalous couplings, the distributions of $p_{W}$, $\cos\theta_{j1}$, $\cos\theta_{j2}$, and
$\phi_{\mathrm{plane}}$ for events with non-zero anomalous couplings were compared with the SM case.  For the comparison we varied two of the
parameters $a,\ b$, and $\tilde{b}$, whilst the third was set to zero.  We then drew probability contours for $\Delta \chi^{2} = 1,\ 2.28$ and
$5.99$, corresponding to $39\%,\ 68\%$ and $95\%$ confidence levels (C.L.) respectively, as in Figs. \ref{fig:cont_a-b} through \ref{fig:cont_b-bt}.

The contour plot in the $a$-$b$ plane (Fig.$\,$\ref{fig:cont_a-b}) shows a linear correlation between $a$ and $b$ due to changes in absolute value
of the $ZH \to \nu\nu WW^{\ast}$ cross section, which increases with $a$ but decreases with increasing $b$.  Note that with our conventions $a \simeq -4.1$ cancels the SM coupling of the Higgs to $W^+W^-$, which means that taking $a \simeq -8.2$ effectively reverses the sign of the SM coupling term.  If we
reverse the sign of the $b$ term in addition, we hence obtain exactly the same distribution provided that the other parameter, $\tilde{b}$, is kept
at zero.  For this reason we observe a second allowed region in Fig.$\,$\ref{fig:cont_a-b}, connected to the first region (containing the SM point
$a=b=0$) by a $180^\circ$ rotation about $(a,b)=(-4.1,0)$.

By the same token, we have two allowed regions for the contours in the $a$-$\tilde{b}$ plane as plotted in Fig.$\,$\ref{fig:cont_a-bt}.
The additional mirror symmetry of the contours about $\tilde{b}=0$ is present because we did not identify  the charge of the charm jets for the $\phi_{\mathrm{plane}}$ measurement.  The prospects for resolving this additional degeneracy by measuring the jet charge are discussed in Section~\ref{sec:cjetchargeid}.

Finally, Fig.$\,$\ref{fig:cont_b-bt} shows the contours in the $b$-$\tilde{b}$ plane for $a=0$.  We observe that these contours are also symmetric under the replacement $\tilde{b}\to -\tilde{b}$, again due to the non-identification of the jet charge.

%
%

\section{Discussions}
\label{sec:discussion}
\subsection{Charge identification of $c$-jets}
\label{sec:cjetchargeid}
If the charge of $c$-jets can be identified, the shape of the $\phi_{\mathrm{plane}}$ distribution will change depending on the sign of $\tilde{b}$. For this reason, it might be possible to identify the sign of $\tilde{b}$ by using the charge identification of $c$-jets. Since the charge identification of $b$- and $c$-jets is considered possible at the ILC by measuring the number of positive and negative charged tracks in the jet clusters, we investigated the sensitivity to the anomalous coupling including identification of the jet charge. 

The efficiency of the charge identification for $c$-jets was assumed to be $14.6\%$ \cite{lcfi} and the same efficiency was used for the backgrounds. We ignored the probability to mis-identify the jet charge as the opposite sign. If we require that the charge of at least one $c$-jet candidate is identified, $27\%$ of $ZH \to \nu \nu W W^{\ast} \to \nu \nu cscs$ events can be selected. By using these events, the $\phi_{\mathrm{plane}}$ distribution was produced considering the relative direction of the positive and negative charge ($\phi_{\mathrm{plane}}^{+-}$). When the charges of both $c$-jet candidates were not identified, $\phi_{\mathrm{plane}}$ was calculated just as the angle between two $c$-jet candidates ($\phi_{\mathrm{plane}}^{00}$). Then, the  sensitivity to the anomalous coupling was evaluated by using $p_{W}$, $\cos\theta_{j1}$, $\cos\theta_{j2}$, $\phi_{\mathrm{plane}}^{+-}$, and $\phi_{\mathrm{plane}}^{00}$. 

\begin{figure}
\centerline{\includegraphics[width=12cm]{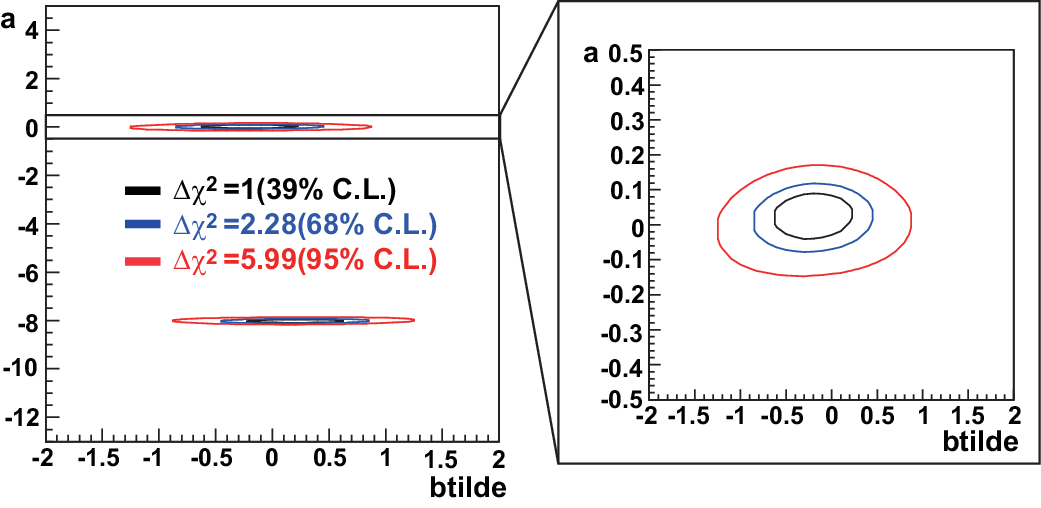}}
\caption{Contours similar to Fig.$\,$\ref{fig:cont_a-bt} plotted in the $a$-$\tilde{b}$ plane. Here, the integrated luminosity is assumed as 1,000 fb$^{-1}$.}
\label{fig:cont_a-bt_cid1000}
\end{figure}

Figure \ref{fig:cont_a-bt_cid1000} shows the probability contours for $\Delta \chi^{2} = 1,\ 2.28,$ and $5.99$ in the $a$-$\tilde{b}$ plane. Here, the integrated luminosity is taken to be 1,000 fb$^{-1}$, since the statistics of $ZH \to \nu \nu WW^{*} \to \nu\nu cscs$ are not sufficient to evaluate the background contamination in the $\phi_{\mathrm{plane}}$ distribution after the charge identification with 250 fb$^{-1}$. In Fig. \ref{fig:cont_a-bt_cid1000}, we can see the weak linear correlation between $a$ and $\tilde{b}$.
The mirror symmetry corresponding to $\tilde{b}\to-\tilde{b}$ is thus broken, although we still have the rotational symmetry corresponding to the combined transformation $(a,\tilde{b})\to(-8.2-a,-\tilde{b})$.

\subsection{Theoretical consideration} 
\label{sec:consideration}

Here we discuss theoretical possibilities 
 to induce the non-renormalizable interactions 
 in Eq.~(\ref{eqn2.1}) as low energy effective theory. 
A simple setup is the Randall-Sundrum model \cite{RS}, 
 in which the gauge hierarchy problem can be solved 
 by virtue of the warped metric. 
In the effective four dimensional theory of this model, 
 the effective cutoff at the TeV scale emerges 
 as $\Lambda = M_P \omega$ with the reduced Planck 
 scale $M_P=2.4 \times 10^{18}$ GeV and 
 the so-called warp factor $\omega \sim 10^{-15}$ 
 associated with the warped metric. 
In this model, we can introduce effective higher-dimensional 
 interactions \cite{FIOY}
\begin{eqnarray} 
 {\cal L}_{\rm int} = 
 \sum_{A=1,2} \; c_A\; \frac{\phi^\dagger \phi}{\Lambda^2} \;
 {\rm tr}\left[
   {\cal F}_A^{\mu \nu}{\cal F}_{A \mu \nu} \right],   
\label{intgauge1} 
\end{eqnarray} 
where $\phi$ is the SM Higgs doublet field, the constants $c_A$ are dimensionless parameters, 
 and ${\cal F}_A$ are the field strengths of the corresponding SM gauge groups, 
 U(1)$_Y$ and SU(2)$_L$. 
After EW symmetry breaking, 
 Eq.~(\ref{intgauge1}) is rewritten as interactions of the Higgs boson with photons, $Z$- and $W$-bosons, 
\begin{eqnarray} 
{\cal L}_{\rm int} &=& 
\frac{c_{WW}}{\Lambda}  \left( \frac{v}{\Lambda} \right) 
 H W^+_{\mu \nu} W^{- \mu \nu} 
+ \frac{c_{ZZ}}{2 \Lambda} 
 H Z^{\mu \nu} Z_{\mu \nu}     \nonumber \\  
&+& \frac{c_{\gamma \gamma}}{\Lambda} \left( \frac{v}{\Lambda} \right) 
   F^{\mu \nu} F_{\mu \nu} 
+ \frac{c_{Z \gamma}}{2 \Lambda} \left( \frac{v}{\Lambda} \right) 
   Z^{\mu \nu} F_{\mu \nu}, 
\label{intgauge2}
\end{eqnarray}
where $W^+_{\mu \nu}$, $Z^{\mu \nu}$ 
 and $F^{\mu \nu}$ are the field strengths 
 of the $W$-boson, $Z$-boson and photon respectively. 
The couplings $c_{WW}$ etc. can be described in terms of the original two couplings 
 $c_1,\ c_2$ and the weak mixing angle $\theta_w$ 
 as follows: 
\begin{eqnarray}
 c_{WW} &=& c_2, \nonumber \\
 c_{ZZ}  &=& c_1 \sin^2\theta_w + c_2 \cos^2\theta_w , 
 \nonumber \\
 c_{\gamma \gamma} &=& c_1 \cos^2\theta_w + c_2 \sin^2\theta_w , 
 \nonumber \\
 c_{Z \gamma} &=& (-c_1+c_2) \sin \theta_w  \cos \theta_w .
\end{eqnarray}
We can identify $ b = c_{WW} v/\Lambda$. 
In the same way, we can obtain $\tilde{b}$ in Eq.~(\ref{eqn2.1}) 
 by effective interactions 
\begin{eqnarray} 
 {\cal L}_{\rm int} = 
 \sum_A \; \tilde{c}_A\; \frac{\phi^\dagger \phi}{\Lambda^2} \;
 \epsilon^{\mu \nu \rho \sigma}
 {\rm tr}\left[
   {\cal F}_{A \mu \nu}{\cal F}_{A \rho \sigma} \right].    
\label{intgauge3} 
\end{eqnarray}  
We can also introduce 
\begin{eqnarray}  
 {\cal L}_{\rm int} = 
  c_0 \; \frac{\phi^\dagger \phi}{\Lambda^2} \;
 \left(D^\mu \phi \right)^\dagger \left(D_\mu \phi \right),  
\end{eqnarray}  
which gives rise to $a=c_0 v/\Lambda$ in Eq.~(\ref{eqn2.1}). 
If we allow $c_A, \tilde{c}_A, c_0 = \pm {\cal O}(10)$, 
 the parameters, $a, b, \tilde{b}$ can be as large as order unity 
 for $\Lambda=1$ TeV.

\section{Summary}
\label{sec:summary}

In this work we have studied the sensitivity of the ILC to dimension-$5$ anomalous Higgs boson couplings to $W^\pm$ pairs, using the decay $H\to WW^\ast$.  For historical reasons, we assumed a Higgs boson mass of $120$~GeV and considered the Higgs boson to be produced in association with a $Z$ boson through the Higgs-strahlung process.  In order to avoid additional combinatorial backgrounds, we required the associated $Z$ boson to decay invisibly into $\nu\bar\nu$ pairs.  A direct measurement of this type is not expected to be possible at the LHC, due to the large QCD background.

Around the SM point $a=b=\tilde{b}=0$, the coefficients of the anomalous couplings can be measured directly at the ILC by examining kinematic distributions such as the on-shell $W$ boson momentum and the angle between $W$ boson decay planes.  Such a measurement will be able to probe the Lorentz structure of the $HW^+W^-$ vertex, providing a direct test of the mechanism of spontaneous symmetry breaking.  The sensitivity may be enhanced by combining the $ZH\to \nu\bar\nu WW^\ast$ channel considered here with other decay modes of $Z$ such as $Z \to \ell^{+} \ell^{-}$ and $Z \to q \bar{q}$, although the large combinatorial background is expected to limit the sensitivity in the case of $Z \to q \bar{q}$.

Although we typically consider the anomalous couplings to be small corrections to the SM term, in principle a SM-like measurement is compatible with two distinct regions around $a=0$ and $a\simeq -8.2$.  This corresponds to a sign change in the coupling constant of the SM $HW^+W^-$ vertex, which is only observable once a sign convention has been fixed by some other coupling (for example, the Yukawa coupling to $b$-quarks).  The sign of the SM term could in principle be measured in interference effects.

\vspace{1.0cm}
\hspace{0.2cm} {\bf Acknowledgments}
\vspace{0.5cm}

The authors would like to thank all the members of the ILD detector
optimization working group \cite{ildoptg} and ILC physics subgroup
\cite{Ref:subgroup} for useful discussions. 
This work is supported in part by the Creative Scientific Research
Grant (No. 18GS0202) of the Japan Society for Promotion of Science, 
JSPS Grant-in-Aid for Scientific Research (No. 22244031 and No. 23000002) 
and the DOE Grants (No. DE-FG02-10ER41714).
RNH is grateful to the Japan Society for the Promotion of Science for support during the initial stages of this project.

\end{document}